\documentclass[12pt,a4paper]{article}

\usepackage{lineno}
\usepackage{authblk}
\usepackage{natbib}
\usepackage{color}
\usepackage{txfonts}
\usepackage{setspace}
\usepackage[dvips]{graphicx}
\usepackage{color}
\usepackage{fancyhdr}

\usepackage{amssymb}

\def\.{\' \i}

\def\B {\mathbf{B}}

\def\U {\mathbf{U}}

\def\u {\mathbf{u}}

\def\b {\mathbf{b}}

\def\n2B {\nabla ^2 \mathbf{B}}

\newcommand{\erf}{{\mathrm{erf}}}

\begin{document}

\title{A mean-field Babcock-Leighton solar dynamo model with long-term variability}

\date{April 2013}

\author[1]{Sabrina Sanchez\thanks{sabrina@on.br}}
\author[2]{Alexandre Fournier}
\author[1]{Katia Pinheiro}
\author[2]{Julien Aubert}
 {\affil[1]{{ \small Observat\'orio Nacional; Rua General Jos\'e Cristino, 77, S\~ao Crist\'ov\~ao, CEP: 20921-400, Rio de Janeiro, RJ- Brasil}}
\affil[2]{{ \small Institut de Physique du Globe de Paris, Sorbonne~Paris~Cit\'e, Univ~Paris~Diderot, UMR~7154~CNRS, F-75005 Paris, France.}}

\maketitle


\section*{Abstract}

Dynamo models relying on the Babcock-Leighton mechanism are successful in reproducing most of the solar 
magnetic field dynamical characteristics. However, considering that such models operate only above
a lower magnetic field threshold, they do not provide an appropriate magnetic field regeneration
process characterizing a self-sustainable dynamo. In this work we consider the existence of an 
additional $\alpha$-effect to the Babcock-Leighton scenario in a mean-field axisymmetric kinematic numerical model. 
Both poloidal field regeneration mechanisms are treated with two different strength-limiting factors. 
Apart from the solar anti-symmetric parity behavior, the main solar features are reproduced: cyclic 
polarity reversals, mid-latitudinal equatorward migration of strong toroidal field, poleward migration 
of polar surface radial fields, and the quadrature phase shift between both. Long-term variability of the 
solutions exhibits lengthy periods of minimum activity followed by posterior recovery, akin to the observed Maunder Minimum. 
Based on the analysis of the residual activity during periods of minimum activity, we suggest that these are caused by 
a predominance of the $\alpha$-effect over the Babcock-Leighton mechanism in regenerating the poloidal field.

\section{Introduction}

	The Sun is a magnetic active star, which undergoes periods of high and low magnetic activity approximately 
each 11 years. Its dynamical behavior imposes important consequences to the terrestrial environment, not 
only producing magnetic storms, which affect satellite operation \citep{baker2000effects}, but also possibly having 
an important role in Earth's climate long-term variability \citep{haigh2003effects}.

 By magnetic activity we understand the appearance of sunspots, characterized by their lower luminosity 
(in comparison with the overall photosphere) and intense magnetic fields \citep{solanki2003sunspots}. Sunspots usually 
appear in pairs of opposite polarities roughly aligned with the E-W direction, as the superficial signature of concentrated azimuthal magnetic fields (toroidal flux tubes), arising from 
the deep of the convection zone by magnetic buoyancy \citep{parker1979sunspots}. It is important to mention that the 
alignment with the equatorial direction is not perfect: sunspot pairs often display a systematic tilt, 
the leading spot being nearer the equator than the following one - Joy's law. Sunspots generally appear 
within a 30$^o$ latitudinal band in each side of the equator, displaying opposite polarity configuration 
in each hemisphere - Hale's polarity law. As the activity cycle is initiated, sunspot appearance migrates 
towards the equatorial region, and after the end of the 11 years cycle they begin again to appear at 
approximately 30$^o$, but with an opposite polarity configuration. This means the full magnetic sunspot 
cycle lasts for twice the activity period. Moreover, sunspots cyclic appearance is directly linked with the 
variability of the large-scale solar magnetic field: during episodes of maximum activity, the polar 
magnetic field undergoes polarity inversion \citep{Makarov2001large}.

All such outstandingly well organized features of the dynamical solar magnetic field 
originate from a natural dynamo process. The magnetic field is regenerated against ohmic dissipation
by electromagnetic induction - convective motions (comprising large-scale, laminar and small-scale  
 turbulent flows) 
produce electric currents which generate in turn secondary magnetic fields 
thereby maintaining 
 the field (see \citet{ossendrijver2003solar} for a review on the subject). These features also indicate that
 the large-scale magnetic field can be decomposed into two main evolving components whose phases are shifted: the toroidal (azimuthal) magnetic field, associated with sunspots, and the poloidal (meridional) 
field, represented by the polar field. Despite the regular character evidenced by the solar cycle, it 
also undergoes amplitude and frequency fluctuations \citep{hathaway2010solar}. The most striking examples of this 
variability are the periods of minimum activity, such as the Maunder Minimum. During this episode, which took place
 from nearly 1645 AD to 1715 AD, sunspots were rarely seen. However, indirect data indicated the persistence, 
although weak, of the solar cycle \citep{beer1998active}. Much discussion exists on the cause of this 
peculiar variability, and how the answers could help access unconstrained properties of the solar 
dynamo mechanism.

Modeling of the solar dynamo has shed light into some of the main processes responsible for the solar 
cycle \citep{charbonneau2010dynamo}. The three processes ($\alpha$-effect, $\Omega$-effect, and 
Bacbcock-Leighton mechanism) discussed in the following are illustrated in Figure~\ref{figure_1}.

\begin{figure}[h]
\includegraphics[width=16.2cm]{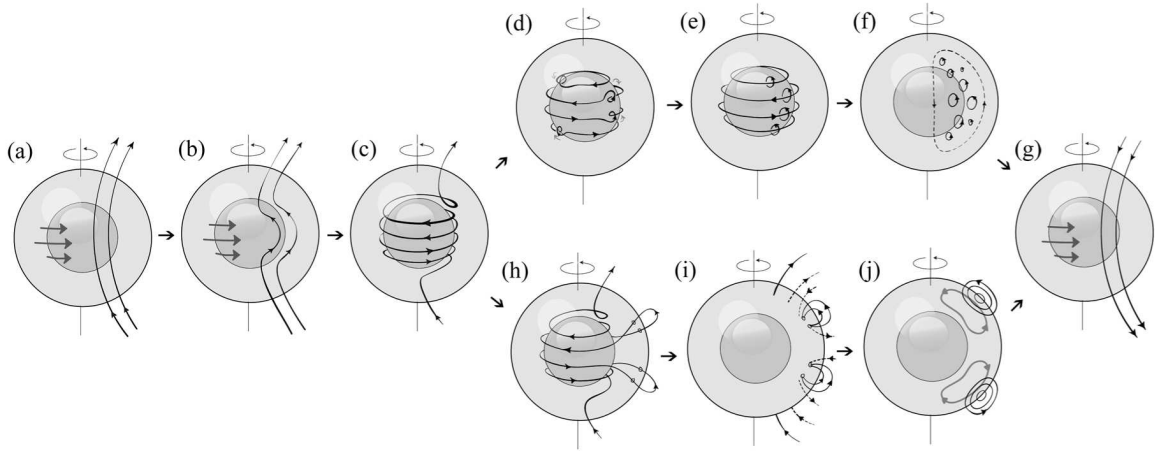}
\caption{Representative scheme of the main processes thought to occur during the solar cycle, departing from (a), an initial poloidal field. (b) and (c) represent the generation of the toroidal field by differential rotation - the $\Omega$-effect. (d) and (e) show the effect of cyclonic turbulence on former toroidal fields, creating small-scale secondary poloidal magnetic fields - the $\alpha$-effect. Averaged, they result in a net electromotive force generating a new large-scale poloidal field (f), closing the first half part of the magnetic cycle with a new poloidal field (g), with opposite polarity than the initial one.  (h) represents the beginning of the Babcock-Leighton mechanism: toroidal flux tubes buoyantly rise to the surface forming sunspots, tilted bipolar regions. In (i), the field from the bipolar region diffuses and reconnects with the correspondents of each hemisphere and with the polar fields. The resulting poloidal flux is advected by meridional circulation to the poles (j), generating the final large-scale poloidal field in (g).}
\label{figure_1}
\end{figure}

The $\Omega$-effect 
symbolizes the shearing action of the differential rotation of the flow on an initial poloidal field, 
giving rise to a toroidal field. Through the advent of helioseismology, the large-scale flow has been 
mapped in detail \citep{schou1998helioseismic}. The region of strongest angular velocity gradients, and so the 
preferred site for the $\Omega$-effect, was found to be at the base of the convection zone: the tachocline 
\citep{howe2000dynamic}. Turbulent motions associated with the Coriolis force, in turn, twist the 
toroidal field, generating a new component of the poloidal field and thus maintaining the solar cycle 
\citep{parker1955hydromagnetic}. This latter process is known as the mean-field $\alpha$-effect, and unlike the 
$\Omega$-effect, is far from being totally understood, as well as the preferred place for its action. It 
is also known that the Lorentz force back-reaction of strong magnetic field inhibits turbulence; so the 
conventional $\alpha$-effect would not lead to great effectiveness in regenerating the poloidal field in
a dynamo within a strong magnetic field regime \citep{cattaneo1996nonlinear}. 
Other mechanisms, like interface dynamos, would overcome this problem, for the $\alpha\Omega$ process 
would occur in the stably stratified layer comprising the tachocline \citep{parker1993solar}.

Alternatively, the Babcock-Leighton mechanism, may have an important role on the poloidal field 
regeneration. Differently from the mean $\alpha$-effect, the Babcock-Leighton mechanism relies 
on the diffusion of the sunspots magnetic field, operating at the solar surface \citep{babcock1961topology,leighton1969magneto}. 
However, the mechanism behind sunspot formation remains elusive (e.g. \citet{guerrero2011dynamo}), 
making it difficult to specify the exact nature of the Babcock-Leighton process in a model. 
In addition, magnetic pumping at the base of the convection zone provides an interesting complement  
to the Babcock-Leighton scenario (e.g. \citet{guerrero2008turbulent}).

Axisymmetric numerical models of the solar dynamo are widely used as a tool to investigate
 the relevance of the main effects supposed to govern the solar cycle \citep{charbonneau2010dynamo}. 
  Many of the solar features 
are well reproduced by numerical models, but no agreement is achieved considering the particular causes. 
The variability of the solar cycle, for example, is generally explained either by stochastic forcing \citep{choudhuri1992stochastic, 
charbonneau2000stochastic} or by dynamical nonlinearities \citep{bushby2006zonal}. Alternatively, simple form of time-delays 
arising from the spatial decoupling of the $\alpha$ and $\Omega$-effects operation place in the solar 
convection zone (like in the Babcock-Leighton case), are known to also yield long-term modulation of the 
solar cycles \citep{wilmot2006time,jouve2010buoyancy}, even reaching a chaotic behavior \citep{charbonneau2005fluctuations}.
In this paper we use a 2D kinematic solar dynamo model merging concepts of the mean-field theory and 
the Babcock-Leighton mechanism, accounting for different magnetic field strength-limiting thresholds, in 
order to achieve solar-like long-term variability.

\section{Model formulation}

To access the variability of the magnetic field in the kinematic context of the solar dynamo, in which 
the flow field is steady and prescribed, the problem reduces to solving the MHD induction equation for 
the magnetic field $B$
\begin{equation}
\label{induction}
\frac{\partial \B}{\partial t}= \nabla \times \left( \U \times \B \right) - \nabla \times 
	\left( \eta_m \nabla \times \B \right),
\end{equation}
where $\U$ is the flow field and $\eta_m$ the magnetic diffusivity. To a first approximation, the 
large-scale magnetic and flow field can be represented as contributions of their large-scale mean 
and small-scale fluctuating parts, $\B=  \langle \B \rangle + \b'$ and $\U = \langle \U \rangle + \u'$, 
respectively. Upon substitution of these quantities in equation (\ref{induction}), and proper averaging, 
we get the mean-field induction equation \citep{moffatt1978field}, given by
\begin{equation}
\label{induction_mf_1}
\frac{\partial \langle \B \rangle}{\partial t}= \nabla \times \left( \langle \U \rangle \times \langle \B \rangle \right) + \nabla \times \left( \langle \mathbf{\u}' \times \mathbf{\b}' \rangle \right) - \nabla \times \eta_m \nabla \times \langle \B \rangle,
\end{equation}
The term $\langle \u' \times \b' \rangle$ corresponds to a mean electromotive force $\mathcal{E}$ arising 
from the interactions of turbulent motions with the small-scale magnetic field. It can thus be written 
in terms of a parameterization of the turbulent effects on the mean-magnetic field as 
$\mathcal{E} = \alpha \B - \beta \nabla \times \langle \B \rangle$. Substitution in the mean-induction 
equation (\ref{induction_mf_1}) gives
\begin{equation}
\label{induction_mf_2}
\frac{\partial \B}{\partial t}= \nabla \times \left( \U \times \B \right)  +  \nabla \times \left( \alpha \B \right) - \nabla \times \eta \nabla \times \B,
\end{equation}
in which the averaging brackets have been omitted, and will remain so throughout. The $\alpha$ 
term represents the turbulent magnetic helicity, due to cyclonic motions oriented by the Coriolis force and 
$\eta= \eta_m + \beta$ is now the effective magnetic diffusivity, covering both magnetic diffusion at 
the microscopic level and turbulent diffusion, respectively.

An additional important issue arises from working within the kinematic context: how should one deal with the 
feedback of the magnetic field on fluid motions (the Lorentz force)? Since the Navier--Stokes  
equation is not solved, this effect ought to be parameterized in the induction equation. This must be done 
in a way which enforces the saturation of the magnetic field, based on the 
equipartition of energy between its small-scale magnetic and turbulent kinetic components. Generally, in the mean-field context, 
the resulting quenching of the magnetic field is heuristically formulated 
as a decreasing function of its intensity, 
\begin{equation}
\alpha(B) = \frac{\alpha_0}{1 + \left( \frac{B}{B_{eq}} \right)^2}, 
\label{quenching_1}
\end{equation}
$B_{eq}$ being the equipartition magnetic field and $\alpha_0$ a typical value characterizing the 
$\alpha$-effect. Studies of magnetoconvection on the solar interior give $B_{eq} \sim 10^4G$ \citep{fan2009magnetic}, 
and lead to conjectures that a turbulent $\alpha$-effect would not represent an effective dynamo mechanism \citep{cattaneo1996nonlinear}.

Alternatively to the conventional $\alpha$-effect, \citet{babcock1961topology} and \citet{leighton1969magneto} suggested 
that the 
diffusion of the magnetic field of sunspot pairs played a crucial role on the process of poloidal field 
regeneration. The bipolar sunspot regions are tilted in a way as the magnetic field of the leading spot, 
nearer the equator, diffuses and reconnects with the field of the leading spot at the other hemisphere, 
which has nearly always an opposite polarity. At the same time, the magnetic field of the following spot 
will also decay and connect with the polar magnetic field, which has opposite polarity. This process
will at some point annihilate the flux in the polar 
region, causing a poloidal polarity reversal. In this case the poloidal 
and toroidal fields regeneration processes are spatially separated (recall Figure~\ref{figure_1}h to
 Figure~\ref{figure_1}j); a mechanism transporting the new polar surface 
magnetic flux generated by the Babcock-Leighton mechanism to the bottom of the convection 
zone is necessary. Facing this requirement, initially a meridional circulation flow was assumed in Babcock-Leighton 
dynamo modeling. As a matter of fact, a poleward meridional flow is indeed observed at the solar surface 
\citep{duvall1979large}. Numerical models comprising this additional physical ingredient are termed flux 
transport dynamos and they are usually successful in reproducing the equatorward tendency of the toroidal 
field \citep{kuker2001circulation}. In summary, the meridional flow not only acts as a means to transport 
the polar surface
magnetic field down to the base of the convection zone where it is transformed into a toroidal field, but it
drags the toroidal field towards the equator as well, yielding a solar-like behavior of the magnetic field.
However, the meridional flow profile used in such model is questionable, 
for it usually consists of a single meridional cell per hemisphere, with a return flow penetrating deep into the 
convection zone. 
From full MHD simulations, the resulting pattern of meridional
circulation is thought to be far more complex, being divided into smaller cells in each hemisphere (e.g. \citet{brun2004global}).

Simulations of the rising of thin magnetic toroidal flux tubes from the base of the convection zone suggest 
that for matching the observed tilts, the flux tubes generating sunspots would need to have an initial magnetic 
field strength ranging from approximately $10^4$ to $10^5$ G. Stronger toroidal flux tubes tend to reach  
the solar surface with almost no tilt, while weaker flux tubes arise at too high latitudes \citep{d1993theoretical}, hence in contradiction with Joy's law. The Babcock-Leighton mechanism should therefore depend 
upon the initial intensity of the toroidal flux tube at the base of the convection zone, just above the tachocline.

Despite the observations supporting the Babcock-Leighton mechanism, much discussion exists on
whether it is the unique responsible for the poloidal field regeneration, an active but not unique 
component of this part of the cycle, or yet a minor contributor to the whole process. The first possibility 
is unlikely, as a dynamo based only on the Babcock-Leighton mechanism is not self-sustainable - it needs 
the formation of sunspots, and therefore operate only above critical toroidal magnetic field buoyancy values. 
A probable scenario is that an additional $\alpha$-effect operates on the convection zone, and the 
resulting magnetic field is the combined product of both poloidal field regeneration processes.

Numerical simulations based on the Babcock-Leighton mechanism have the tendency 
to produce equatorially symmetric solutions, in opposition to what is observed in the Sun (Chatterjee et al. 2004). Recent results have 
shown that an $\alpha$-effect located within a thin layer just above the tachocline is more successful at
 yielding equatorially antisymmetric solutions \citep{bonanno2002parity,dikpati2001flux}. 
In this study, we focus on this scenario for the location of the $\alpha$-effect (although other options
 are possible, see for instance \cite{kapyla2009alpha}.). 
Residual turbulent 
motions acting on toroidal flux tubes right above the tachocline, before these reach a critical
 strength and become buoyant, support such a disposition 
 of the $\alpha$-effect. This concept is applied in the present 
model, in addition to the Babcock-Leighton effect at the solar surface.

\section{Mathematical description}

Under the assumption of axisymmetry, the magnetic and flow fields are written in terms of their poloidal 
and toroidal components in spherical coordinates $(r,\theta,\phi)$ as       
\begin{eqnarray}
\B(r,\theta,t) &=& \nabla \times  \left[ A_{\phi}(r,\theta,t) \mathbf{\hat{e}_{\phi}}  \right] 
	+ B_{\phi}(r,\theta,t) \mathbf{{\hat{e}}_{\phi}}, \label{B_pol_tor} \\
 \U(r,\theta) &=& \mathbf{\u_{p}}(r,\theta) + r \sin\theta  \, \Omega(r,\theta) \mathbf{{\hat{e}}_{\phi}},
\label{V_pol_tor}
\end{eqnarray}
in which $A_{\phi}$ is the poloidal potential and $B_{\phi}$ is the toroidal field. The steady flow 
profile $\U$ is given by the meridional circulation $\u_p$ and the differential rotation $\Omega$. The poloidal-toroidal 
decomposition of the magnetic field enables a separation of the mean induction equation (\ref{induction_mf_2}) 
into two partial differential equations for $A_{\phi}$ and $B_{\phi}$, 
\begin{eqnarray}
\frac{\partial A_{\phi}}{\partial t} + \frac{Rm}{r \sin\theta} \mathbf{u_p} \cdot \nabla \left( r \sin\theta A_{\phi} \right) &=&  \tilde{\eta}_p \left( \nabla^2 - \frac{1}{r^2 \sin\theta^2} \right) A_{\phi} \nonumber \\ 
&+& C_{\alpha} \alpha (r,\theta; B_{\phi}) B_{\phi} + C_S S(r,\theta;B_{\phi}^{tc}) B_{\phi}^{tc},
\label{induction_pol} \\
\nonumber \\
\frac{\partial B_{\phi}}{\partial t} + Rm \; r\sin\theta \nabla \cdot \left( \frac{\mathbf{u_p} 
	B_{\phi}}{r \sin\theta}  \right) &=& \tilde{\eta}_t \left( \nabla^2 - \frac{1}{r^2 \; \sin \theta^2} \right) B_{\phi} \nonumber\\
&+& \frac{1}{r} \frac{\partial \tilde{\eta}_t}{\partial r} \frac{\partial (r B_{\phi})}{\partial r} + C_{\Omega} \; r \; 
	\sin\theta \left( \nabla \times A_{\phi}\mathbf{\hat{e}_{\phi}} \right) \cdot \left( \nabla \Omega \right).
    \label{induction_tor}
\end{eqnarray}

The appearance of the three numbers quantifying the strength of the processes discussed in the introduction, namely  
\begin{eqnarray*}
C_{\Omega}&=&\Omega_{eq}R^2/\eta_{s},\\ 
C_{\alpha}&=&\alpha_0R/\eta_s,       \\
C_S&=&S_oR/\eta_s, 
\end{eqnarray*}
and of the magnetic Reynolds number, 
\begin{eqnarray*}
Rm&=&u_oR/\eta_s,  
\end{eqnarray*}
results from the nondimensionalization of the equations using the solar radius $R$ 
as the characteristic length scale and the effective magnetic diffusion time $R^2/\eta_s$ as the characteristic 
time scale. $B_{\phi}^{tc}$ is the toroidal field just above the tachocline; $\Omega_{eq}$, $\alpha_o$, $S_o$ 
and $u_o$ are the rotation rate at the equator, the typical magnitudes of the poloidal source terms, 
 and the peak velocity 
of the meridional flow at the surface. $\tilde{\eta}_p$ and $\tilde{\eta}_t$ are the normalized effective 
magnetic diffusivities for the poloidal and toroidal components, respectively. Note that the extra $\alpha$-term 
that would arise in equation (\ref{induction_pol}) has been neglected, as it is common in the solar case 
approximation to suppose that $C_{\Omega}\gg C_{\alpha}$. The flow specifications, $u_p$ and $\Omega$ are the 
same as the ones used in the reference work of Dikpati and Charbonneau (1999), and are shown in the left panels 
of Figure \ref{figure_2}. In this paper, the tachocline will comprise the region from the top of the radiative zone
$r_r=0.6R$, to the top of the region with the largest radial angular velocity gradients, $r_{tc}=0.7R$.  

\begin{figure}
\begin{center}
\includegraphics[width=8cm]{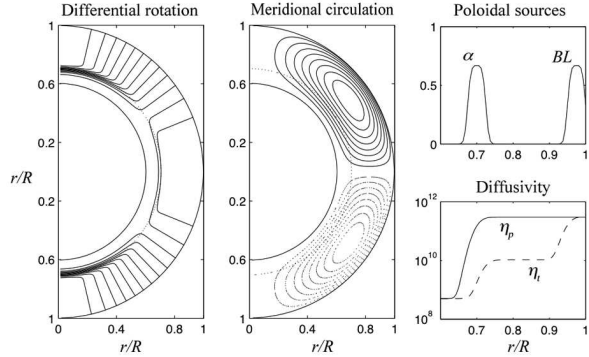}
\caption{Specifications of the model. Left: isocontours of $\Omega(r,\theta)$  based on analytic fit of the differential rotation profile from helioseismology data. Center: meridional circulation streamlines (full line - counterclockwise flow, dotted line - clockwise flow). Right top: $\alpha$ and Babcock-Leighton poloidal source radial profiles. Right bottom: effective magnetic poloidal (continuous line) and toroidal (dashed line) diffusivity radial profiles.}
\label{figure_2}
\end{center}
\end{figure}

It is worth noticing that the $S$ term, representing the Babcock-Leighton poloidal regeneration process 
at the surface, has been added in an ad-hoc manner to equation (\ref{induction_pol}). Unlike the 
$\alpha$-term, it is non-local - it depends on the toroidal field at the tachocline. This comes from the 
latitudinal tilt given by Joy's law being rather dependent on the initial magnetic field strength of the 
rising flux tube, therefore at the tachocline \citep{d1993theoretical}. 

The $\alpha$-effect in equation (\ref{induction_pol}) must be suppressed for magnetic field strengths above 
a limiting value associated with energy equipartition. Therefore, the $\alpha$-term will be written following 
equation~(\ref{quenching_1}), as
\begin{equation}
\label{quenching_2}
\alpha (r,\theta;B_{\phi}) = \frac{1}{1+\left(\frac{B_{\phi}}{B_{eq}}\right)^2} f_{\alpha}(r,\theta),
\end{equation}
where $B_{eq} = 10^4$ G and $f_{\alpha}(r,\theta)$ is a function of spatial coordinates given by
\begin{equation}
f_{\alpha}(r,\theta) = \frac{1}{4} \left[ 1 + \erf \left( \frac{r-r_1}{d_1} \right) \right] 
	\left[ 1 - \erf \left( \frac{r-r_2}{d_2} \right) \right] \cos \theta \sin \theta,
\end{equation}
where $r_1 = 0.675 R$, $r_2 = 0.725 R$ and $d_1=d_2=0.01 R$. This function constrains the $\alpha$-effect to a
thin layer at the base of the convection zone, just above the tachocline, and to mid-latitudes. As mentionned above, other options are possible, but the detailed exploration of these is beyond the scope of the present study. (In passing,  we tried the end-member case of a quasi-homogeneous distribution of $\alpha$, which 
can produce a solar-like dynamo, but over a limited range of $C_{\alpha}$, $ 1 \lesssim C_{\alpha} \lesssim 3 $.) 
Similar conjectures apply to the Babcock-Leighton $S$ source term in equation~(\ref{induction_pol}), with the 
difference that it operates between lower and upper limiting values, 
\begin{equation}
S(r,\theta;B_{\phi}^{tc}) = \frac{1}{4} \left[ 1 + \erf \left( B_{\phi}^{tc \; 2} - B_{\phi \; min}^{tc \; 2} \right) \right] 
	\left[ 1 - \erf \left( B_{\phi}^{tc \; 2} -  B_{\phi \; max}^{tc \; 2} \right) \right] f_S(r,\theta).
\end{equation}
Here $B_{\phi \; min}^{tc} = 10^4$ G, $B_{\phi \; max}^{tc} = 10^5$ G and the radial and latitudinal 
distribution $f_S(r,\theta)$ is given by
\begin{equation}
f_S(r,\theta) = \frac{1}{4} \left[ 1 + \erf \left(\frac{r-r_3}{d_3} \right) \right] \left[ 1 + \erf\left( 
	\frac{r-r_4}{d_4}\right)  \right] \cos \theta \sin \theta ,
\end{equation}
where $r_3 = 0.95R$, $r_4 = 1.0 R$ and $d_3 = d_4 = 0.01 R$, restricting the Babcock-Leighton mechanism to 
the near-surface layers. The radial profiles of the $\alpha$ and $S$ poloidal source terms are shown in the top right panel of Figure~\ref{figure_2}.

The effective diffusivity follows the concept of \cite{chatterjee2004full}, who
parameterize the suppression of turbulent diffusion by separating the diffusivity into a poloidal component  
$\eta_p$ and a toroidal component $\eta_t$.
\begin{equation}
\eta_p = \eta_r + \eta_s \; \frac{1}{2} \left[ 1 + \erf \left(\frac{r-r_5}{d_5} \right) \right],
\end{equation}
\begin{equation}
\eta_t = \eta_r + \eta_{cz} \; \frac{1}{2} \left[ 1 + \erf \left(\frac{r-r_6}{d_6} \right) \right] + \eta_s \; 
	\frac{1}{2} \left[ 1 + \erf \left( \frac{r-r_7}{d_7} \right) \right],
\end{equation}
in which $r_5=0.7 R$, $r_6=0.72 R$, $r_7=0.95 R$ and $d_5=d_6=d_7=0.025 R$. Such separation comes from the 
fact that the toroidal field, at least in the deeper layers, tends to be much more intense than the poloidal field, 
and therefore more effective in suppressing turbulent diffusion. $\eta_r$ is the diffusivity 
near the radiative zone, $\eta_{cz}$ the diffusivity in the turbulent convection zone associated with the 
toroidal magnetic field, and $\eta_s$ the diffusivity in the radial surface layers and for the weaker poloidal 
field. They are set in this study to $\eta_r = 5 \times 10^8$ cm$^2$/s, $\eta_{cz}=1 \times 10^{10}$~cm$^2$/s 
and $\eta_s = 3 \times 10^{11}$~cm$^2$/s. The corresponding radial profiles are shown in the bottom right panel of Figure~\ref{figure_2}.

\section{Numerical simulation}

Given the physical ingredients described in the previous section, equations~(\ref{induction_pol}) and~(\ref{induction_tor}) are solved 
numerically over a regular grid in an annular meridional plane covering $\theta \in [0,\pi]$ and $r \in [0.6,1]$, 
that is, from slightly below the tachocline up to the solar surface. The boundary and initial conditions  
are the following: the inner boundary condition is set as to represent the radiative 
zone as a perfect conductor. An approximation of this condition gives that at $r_r = 0.6$, $A_{\phi}=B_{\phi}=0$ 
 \citep{chatterjee2004full}. The outer boundary condition is that of a vacuum region, requesting the magnetic field to connect with a 
potential field in the exterior region \citep{dikpati1999babcock}. As for the initial condition, we use 
a dipolar field permeating the convective envelope. In this case, $A_{\phi} = \sin \theta /r^2$ for $R \geq r \geq 0.7 R$ 
and zero elsewhere, whereas $B_{\phi} = 0$ everywhere.

The solution procedure was performed by an adaptation of the Parody code \citep{dormy1997modelisation,dormy1998mhd,aubert2008magnetic}, based on a 
pseudo-spectral method. It rests on a spherical harmonic expansion of the angular dependence of the poloidal 
and toroidal scalars and finite differences in the radial direction. More details 
on the code, including benchmarks with published numerical solutions \citep{jouve2008solar}, are presented in the Appendix. The results presented in the following were obtained with 
 spectral truncation $L_{max} = 65$, number of radial points $N_r=65$ and a constant, non-dimensional time stepping size  $\Delta t = 5 \times 10^{-6}$.

\section{Results}

On the basis of helioseismic data ($\Omega_{eq} = 2 \pi \times 460.7$ nHz), we fix $C_{\Omega} = 4.7 \times 10^4$; 
in this case, variations of the free parameters $C_{\alpha}$, $C_S$ and $Rm$ allow for a broad range of solutions. 
Meridional circulation measured at the solar surface at mid-latitudes displays an average value of $15$ m/s. 
Considering $Rm$ varying around this value, from $318$ to $378$, the minimum configuration for a proper dynamo 
solution consists of $ C_{\alpha} \simeq 2$ and $C_S \simeq 0.5$. In order to access the solar representativeness 
of the solution, some observable aspects should be matched (Charbonneau 2010):

\begin{enumerate}
\item Cyclic polarity reversals with approximately 11 years periodicity;
\item Strong deep toroidal fields ($ \sim 10^4-10^5 $ G) at a 30$^o$ latitudinal belt migrating equatorward;
\item Poleward migration of the polar radial field ($ \sim 10-100$~G);
\item Phase lag of $\pi/2$ between the deep mid-latitudinal toroidal and surface polar fields;
\item Antisymmetric coupling of the magnetic fields between the hemispheres;
\item Long-term variability of the solar cycle.
\end{enumerate}

A useful way to analyze the solar semblance of simulated results is to display the magnetic field in a 
time-latitude map and compare it with proper synoptic magnetograms and sunspot butterfly diagrams \citep{hathaway2010solar}. 
A reference solution of the model is displayed in this form in Figure \ref{figure_3}, in which the surface 
polar field contours are superimposed with the gray-scale map of the toroidal field just above the tachocline 
($r_{tc} = 0.7 R$). This case was chosen because it met most of the obserational requirements listed above. Criterion 1, for example: 
the periodicity associated with the solar cycle is of 10.95 years. The main dependence of the periodicity resides 
on the strength of meridional circulation, a well-known characteristic of flux transport dynamos \citep{dikpati1999babcock}. Using a similar model as the present one, \citet{charbonneau2005fluctuations} observed the persistent 
key part played by the meridional circulation in setting the cycle periodicity.

The magnetic field morphology presented in criteria 2 and 3 is also achieved, even though the polar strength 
of most of the solutions (peaking at $1,700$ G of the reference case in Figure~\ref{figure_3}) is an order of magnitude higher than the observed one.  
A phase lag of approximately $\pi/2$ between poloidal and toroidal fields cited in criterion 4 is also a general 
property of flux transport Babcock-Leighton models \citep{dikpati1999babcock}.

\begin{figure}
\begin{center}
\includegraphics[width=16.2cm]{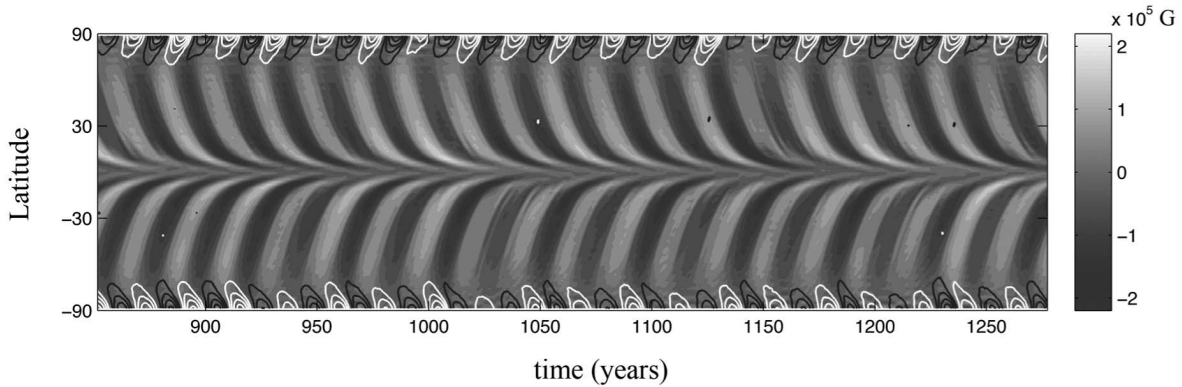}
\end{center}
\caption{Butterfly diagram of the dynamo solution corresponding to $C_S = 1.0$, $C_{\alpha} = 8.0$ and $Rm = 318$. The gray scale map represents the toroidal field at the tachocline, whereas high latitude black and white contours represent the radial field at the solar surface, with a maximum value of 1,700 G.}
\label{figure_3}
\end{figure}

On the other hand, the parity requirement 5 is still an issue. Although accounting for an $\alpha$-effect at a 
thin layer above the tachocline was thought to help yielding anti-symmetric solutions \citep{dikpati2001flux,bonanno2002parity}, the parity coupling was not straightforward in this model case. Actually, there does 
not seem to exist a clear preferred mode for the solutions: they vary between periods of symmetric, anti-symmetric 
and out of phase modes. In the reference case, as it is noticeable in Figure \ref{figure_3}, the activities within 
each hemisphere are slightly out of phase. This probably originates from the chaotic nature of the solutions. 
For higher $C_S$ and $C_{\alpha}$, there is a tendency for the magnetic fields to evolve
independently in each hemisphere, with no stable phase lag. 

\citet{charbonneau2005fluctuations} analyzed the general 
chaotic behavior in a Babcock-Leighton dynamo and ascribed its cause to time-delays connected with 
the  spatial segregation of the toroidal and poloidal field regeneration processes. 
Long-term variability (recall item 6 above) is a consequence of a chaotic behavior. Figure \ref{figure_4} 
displays the evolution of the toroidal magnetic field energy at a certain latitude (toroidal magnetic field 
energy is generally used as a proxy for activity cycle amplitude). We observe frequent short periods of minimum activity with a duration of approximately 3 solar cycles -- we typically get 8 of these every 1,000 yr.  
Moreover, the model also reveals periods of extended minimum activity, reminiscent of the Maunder Minimum, 
in which the cycle is apparently not fully developed, but persists with a kind of residual activity, lasting for approximately 500 years.

\begin{figure}
\begin{center}
\includegraphics[width=16.2cm]{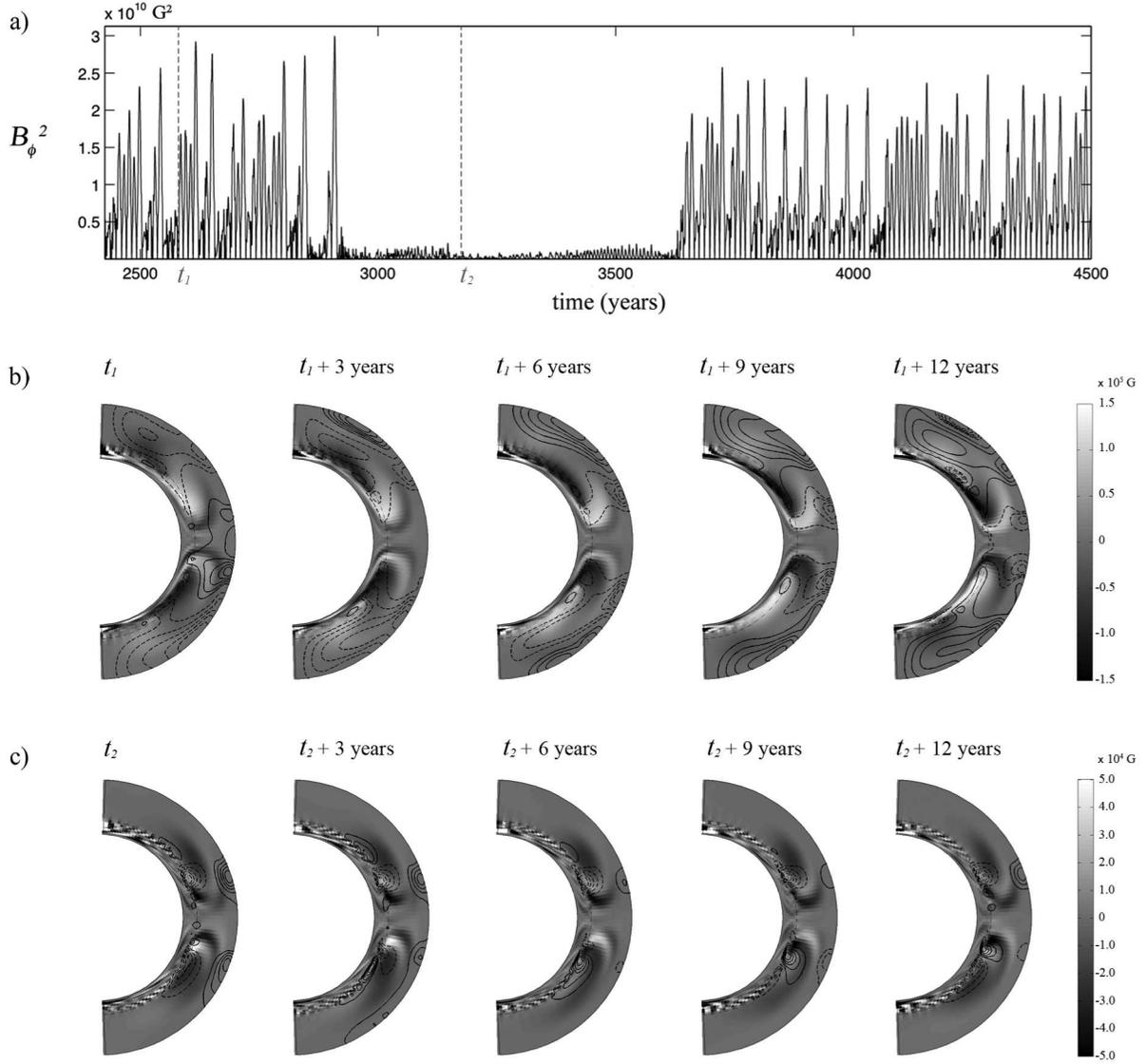}
\end{center}
\caption{(a) Time series of the toroidal magnetic energy at the tachocline at 20$^o$ latitude for the reference solution. In addition to the clear modulation of the solar cycle note the presence of an extended period of minimum activity. (b) and (c) show the toroidal magnetic field (gray scale map) and poloidal potential (contours) on a meridional plane at different times (during a normal phase and a quiescent phase, respectively).}
\label{figure_4}
\end{figure}

The residual activity episode in Figure \ref{figure_4}a suggests that there are two different regimes for the 
large-scale magnetic field behavior. In the
same figure, we show the evolution of the spatial distribution of the poloidal and
toroidal magnetic fields in the meridional plane at different epochs, one during
regular activity (Figure \ref{figure_4}b) and another during the episode of solar quiescence  
(Figure \ref{figure_4}c). During the normal activity period, the magnetic field is mainly large-scale. 
At the tachocline, a strong
toroidal field is created by the shearing effect of differential rotation, from this toroidal field,
a poloidal field is created at the solar surface by means of the Babcock-Leighton mechanism. The role of the
meridional circulation in setting the timing of the solar cycle is clear: the flow transports the toroidal field
at the botton of the convection zone towards the equator, which generates the solar-like orientation of the butterfly
diagram (recall Figure~\ref{figure_3}). On the other hand, 
during the quiescent period (recall Figure~\ref{figure_4}c),
 the magnetic field is small-scale and the solar cycle is confined to the bottom of the convection zone. These differences
points to a drastic change in the underlying dynamo mechanism. 
To investigate this further, we now analyze separatedly the $\alpha$-effect and the Babcock-Leighton mechanism in two distinct dynamo models.

Figure \ref{figure_5} shows the toroidal magnetic energy time series of those two models. 
The poloidal and toroidal fields during half a
solar cycle are also shown in the meridional planes. In the Babcock-Leighton case (Figure~\ref{figure_5}a), the solar cycle 
evolution is smooth and tends to display a persistent weak-strong amplitude bundling configuration (therefore 
with twice the cycle period). This feature, resembling the observed solar pattern known as the Gnevyshev-Ohl rule, is also 
a consequence of the time delays inherent to the Babcock-Leighton mechanism  \citep{charbonneau2007fluctuations}.
The meridional plots of the magnetic field in Figure \ref{figure_5}b show that {$\bf B$} is large scale, which matches the overall description of the normal activity period of the reference model, depicted in Figure \ref{figure_4}b. In the
case of a pure Babcock-Leighton scenario, the toroidal field shows a moderate level of antisymmetry about the equator. 

The situation is different in the pure $\alpha$-effect case,
shown in Figures \ref{figure_5}c and \ref{figure_5}d: the typical amplitude of any given
cycle is much lower than in the pure Babcock-Leighton scenario, the cycle period 
is about twice as long, and there 
are additional high frequency oscillations superimposed to the solar cycle.
Figure \ref{figure_5}d shows that in contrast with the situation of a 
Babcock-Leighton dynamo, the field is mostly small-scale, in agreement with the
appearance of higher frequencies in the $\alpha$-effect model time-series (see the power spectra in \ref{figure_5}c). 

\begin{figure}
\begin{center}
\includegraphics[width=16.2cm]{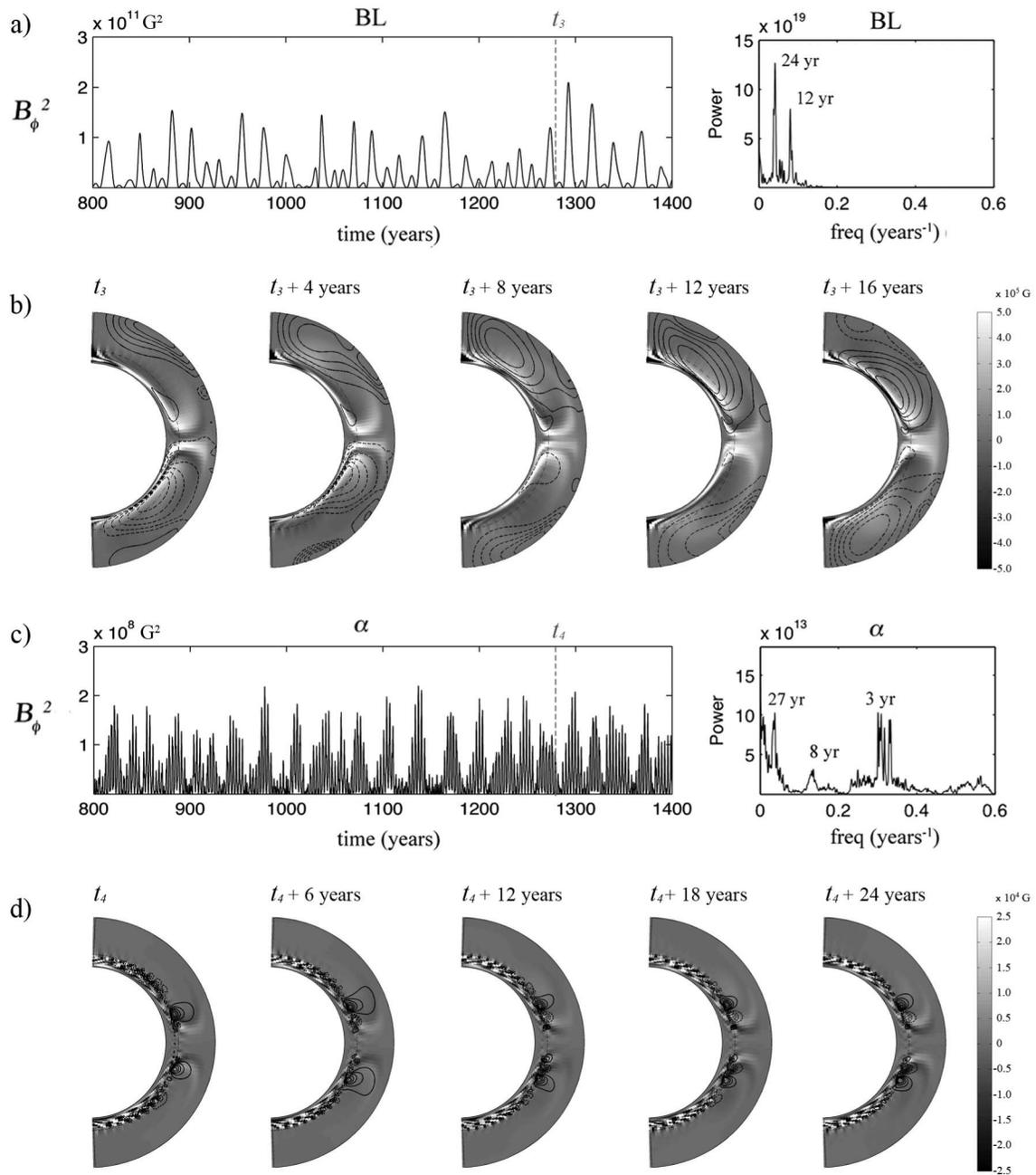}
\end{center}
\caption{Characteristic behavior of the dynamo with (a) only Babcock-Leighton poloidal 
source ($C_S = 4.0$) at the surface and (c) $\alpha$-effect ($C_{\alpha} = 8.0$) at the tachocline, 
for $Rm = 378$, showing the toroidal magnetic energy at the tachocline at 20$^o$ latitude and its power spectra. Similarly as in Figure \ref{figure_4}, (b) and (d) show the magnetic field at a meridional plane for each case.}
\label{figure_5}
\end{figure}

Aware of the different behaviors concerning the different poloidal field regeneration processes, it is 
possible that the period of minimum activity involves the preponderance of the tachocline $\alpha$-effect 
over the Babcock-Leighton mechanism. In such a case, lower initial poloidal fields would result in the generation 
of toroidal field below the buoyancy instability limit, leading to few sunspot formations. The long-term solar
cycle variability would then be mainly driven by the tachocline $\alpha$-effect, generating weak toroidal 
fields during periods of minumum activity, but being able to restart the Babcock-Leighton mechanism when the upper
threshold of toroidal field strength is achieved. This situation is reminiscent of observed minimum activity periods, such as the Maunder Minimum. In addition, note that this long-lasting quiescent situation is rather rare in our model: 
only twice did we observe quiescent periods lasting for more than 500 years in our 20,000 year long integration (short-lived quiescent periods are more frequent, see above). It is also worth mentioning that because of the decoupling of the magnetic field between hemispheres, the minimum activity episode of Figure 4a neither starts nor ends simultaneously in the North and in the South. In this case, the southern hemisphere enters the minimum activity phase approximately 200 years after the northern hemisphere.

Most of the mean-field kinematic solar
dynamo simulations able to reproduce the minimum activity
periods rely either on the introduction of stochastic forcings
 \citep{charbonneau2005maunder}, or on the somehow arbitrary manipulation of the meridional flow and/or Babcock-Leighton poloidal sources (e.g. \citet{karak2010importance}). Here, on the account of the results we presented, we may argue that the $\alpha$-effect located at the tachocline effectively replaces the stochastic forcing in producing long-lasting phases of minimum activity.

\section{Summary and Conclusion}

In view of the not self-sustainable character of a solar dynamo relying only on the Babcock-Leighton mechanism 
for poloidal field regeneration, we have considered an additional $\alpha$-effect operating in a thin layer 
above the tachocline, originating from the turbulent effects on magnetic flux tubes just above a critical 
buoyancy level. Accounting for different limiting ranges of magnetic field on the operation of each effect, 
concerning their different natures, we have obtained a dynamo solution reproducing the basic solar magnetic 
field dynamic features, namely: cyclic reversals with a $\sim$ 11 year periodicity, equatorward migration of 
the activity belt, poleward migration of the polar radial field, proper phase lag between both, and long-term 
variability resembling the solar one.

Appropriate antisymmetric magnetic coupling between the hemispheres remains an issue,
 even if the location of the $\alpha$-effect at the tachocline had been suggested as a way to solve the parity problem. In fact, 
the hemispheres appear to behave in a rather dissociated way, not showing any preferred relaxation mode. The decoupling may originate from the chaotic nature of the solution, making the magnetic field ${\bf B}$ evolve independently in each hemisphere. Further investigations on the parity topic are needed, possibly relying 
upon the joint spherical harmonic analysis of the modelled ${\bf B}$ and that of
 the  ${\bf B}$ observed at the surface of the Sun (see e.g.  
  \citet{stenflo1986global} and \cite{derosa2012solar}).

Our study spontaneously 
presents a Maunder-like grand minimum which, in comparison 
 with other studies based on similar mean-field kinematic dynamo models, does not require the addition of
 a stochastic forcing to the right-hand side of the dynamo equations. 
We conclude by suggesting that grand minima periods could be 
caused by an intermittent phase of the solar dynamo, during which the sole  
deep and weak $\alpha$-effect is 
responsible for the regeneration of the poloidal field. Why the transition 
 from a Backbock-Leighton dominated regime to an $\alpha$-effect dominated
 regime occurs remains a matter of investigation.

\section{Acknowledgements}
Sabrina Sanchez would like to express her gratitude to Dr.~Oscar Matsuura for his valuable suggestions and fruitful discussions. We thank A. S. Brun for discussions and comments on an earlier version of the manuscript. 
This work is a result of a M.Sc. project in collaboration between Observat\'orio Nacional and IPGP. We thank both institutions for the available facilities and the Brazilian agency CAPES for funding this research. The contribution of AF and JA is IPGP contribution number $NNNN$. 

\newpage
\section{Appendix}

\begin{center}
{\large \textbf{Parody Code - Mean Field Benchmarking}}
\end{center}

The Parody code used in this work was originally proposed for full 3D MHD dynamo simulations (ACD code, 
benchmarked in \citet{christensen2001numerical}, see \cite{dormy1998mhd} and \cite{aubert2008magnetic}). 
The magnetic field is decomposed according to
\begin{equation}
\B = \nabla \times \nabla \times ( B_p \; \mathbf{r} ) + \nabla \times ( B_t \; \mathbf{r} ),
\end{equation}
and the scalar potentials further expanded on a spherical harmonic basis
\begin{equation}
B_{p,t} = \sum_{\ell = 1}^{L_{max}} \sum_{m=0}^{M_{max}} B_{p,t \;\ell}^{\; \; \; \; m}(r,t) \; Y_{\ell}^{m} (\theta,\phi).
\end{equation}
The radial dependence is treated by a second-order finite-differencing scheme. Time integration uses a 
Crank-Nicholson scheme for diffusion terms and an Adams-Bashforth scheme of order 2 for the nonlinear terms. The actual 
equations to be solved are the radial curl and radial curl of the curled induction equation (\ref{induction}). 
Further adaptation to the axisymmetric ($M_{max}=0$) mean-field scenario included a change of the magnetic 
 boundary conditon at the inner 
boundary, the specification of a steady flow in terms of a differential rotation and a meridional circulation,  the construction of the 
proper diffusivity profiles and the addition of the $\alpha$ and $S$ source terms to the right-hand side of the induction equation. The perfectly conducting inner boundary condition implies setting $B_p = 0$ and $\partial (r B_t)/\partial r = 0$ 
at the inner boundary.

The code was tested by comparing its predictions with the published reference solutions of a community mean-field 
 benchmark \citep{jouve2008solar}. The goal here is to compute critical 
dynamo numbers and cycle frequencies for different dynamo models, involving either an $\alpha \Omega$ scenario (cases A and B, differing only in 
the prescribed diffusivity) or a Babcock-Leighton scenario (case C). 
Table \ref{table_1} displays the values obtained with our code against the reference ones. The butterfly diagrams for the 
supercritical cases SB and SC (which incorporate an $\alpha$-quenching) are displayed in Figure \ref{figure_6}.

\begin{table}[b]
\begin{center}
\small
\caption{Comparison of the critical dynamo numbers $C_{\alpha,S}^{crit}$ and
frequency of the solar cycle $\omega$
within the benchmark cases A, B and C from \citet{jouve2008solar}. The spatial
and temporal resolutions are given in terms of radial points and harmonic degree truncation 
($N_r \times L_{max}$) and time-step size $\Delta t$.
}
\begin{tabular} {c|cc|cc|cc}

\hline \hline

 & \multicolumn{4}{c|}{Results} & \multicolumn{2}{c}{Reference} \\
 
 \cline{2-7}
 
 Case & Resolution & $\Delta t$ & $C_{\alpha,s}^{crit}$ & $\omega$ & $C_{\alpha,s}^{crit}$ & $\omega$  \\
\hline

A    & 71 $\times$ 71   & $5\times10^{-5}$ & 0.358 & 158.00 & 0.387 $\pm$ 0.002 & 158.1 $\pm$ 1.472 \\
B    & 71 $\times$ 71   & $5\times10^{-5}$ & 0.406 & 172.01 & 0.408 $\pm$ 0.003 & 172.0 $\pm$ 0.632 \\
C    & 120 $\times$ 120 & $ 1\times10^{-6}$ & 2.545 & 534.6 & 2.489 $\pm$ 0.075 & 536.6 $\pm$ 8.295 \\

\hline
\end{tabular}
\label{table_1}
\end{center}
\end{table}

\begin{figure}
\begin{center}
\includegraphics[width=8cm]{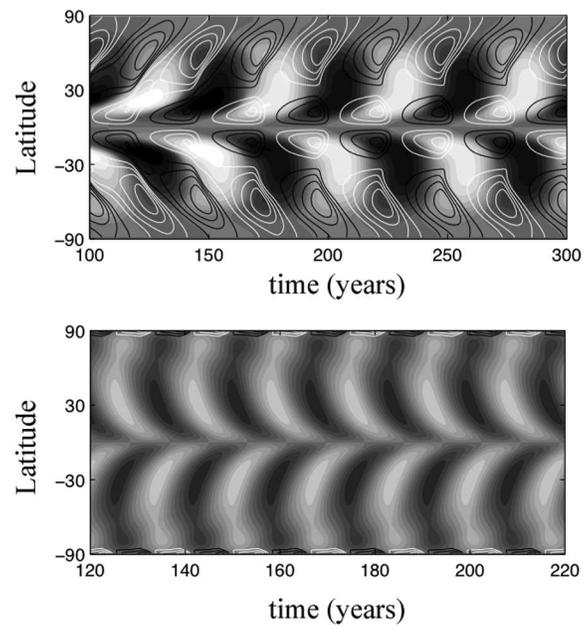}
\end{center}
\caption{Butterfly diagram summarizing the Benchmark cases to be compared with \citet{jouve2008solar} Figures 9 and 14 respectivelly. 
Contours refer to the radial 
field at the surface an gray scale map to the toroidal field at the tachocline. Upper panel: 
$\alpha\Omega $ dynamo from case SB. Lower panel: Babcock-Leighton dynamo from case SC.}
\label{figure_6}
\end{figure}

\newpage
\bibliography{biblio_alpha,extra_AABC}

\bibliographystyle{apalike}

 \end{document}